\documentclass{aastex631}

\usepackage{float}

\usepackage{amsmath}
\usepackage{booktabs}
\usepackage{multirow}
\begin{document}

\title{Spatially resolved polarization variation of the Crab Nebula}

\author[0009-0007-5244-2379]{Chao Zuo}
\affiliation{Guangxi Key Laboratory for Relativistic Astrophysics, School of Physical Science and Technology, Guangxi University, Nanning 530004, China}

\author[0000-0002-0105-5826]{Fei Xie}
\correspondingauthor{Fei Xie}
\email{xief@gxu.edu.cn}
\affiliation{Guangxi Key Laboratory for Relativistic Astrophysics, School of Physical Science and Technology, Guangxi University, Nanning 530004, China}
\affiliation{INAF Istituto di Astrofisica e Planetologia Spaziali, Via del Fosso del Cavaliere 100, 00133 Roma, Italy}

\author[0000-0002-3776-4536]{Mingyu Ge}
\affiliation{Key Laboratory for Particle Astrophysics, Institute of High Energy Physics, Chinese Academy of Sciences, Beijing 100049, China}

\author[0000-0002-9370-4079]{Wei Deng}
\affiliation{Guangxi Key Laboratory for Relativistic Astrophysics, School of Physical Science and Technology, Guangxi University, Nanning 530004, China}

\author[0009-0007-8686-9012]{Kuan Liu}
\affiliation{Guangxi Key Laboratory for Relativistic Astrophysics, School of Physical Science and Technology, Guangxi University, Nanning 530004, China}

\author[0000-0001-8916-4156]{Fabio {La Monaca}}
\affiliation{INAF Istituto di Astrofisica e Planetologia Spaziali, Via del Fosso del Cavaliere 100, 00133 Roma, Italy}
\affiliation{Dipartimento di Fisica, Università degli Studi di Roma “Tor Vergata,” Via della Ricerca Scientifica 1, 00133 Roma, Italy}

\author[0000-0003-0331-3259]{Alessandro {Di Marco}}
\affiliation{INAF Istituto di Astrofisica e Planetologia Spaziali, Via del Fosso del Cavaliere 100, 00133 Roma, Italy}

\author[0009-0000-2414-9449]{Wenhao Wei}
\affiliation{Guangxi Key Laboratory for Relativistic Astrophysics, School of Physical Science and Technology, Guangxi University, Nanning 530004, China}

\author[0000-0002-5965-7432]{Wei Chen}
\affiliation{Guangxi Key Laboratory for Relativistic Astrophysics, School of Physical Science and Technology, Guangxi University, Nanning 530004, China}

\begin{abstract}

We examined the spatially resolved polarization variations in the Crab Nebula over two years, using observational data from the Imaging X-ray Polarimetry Explorer (IXPE), and offer key insights into its magnetic field structures and evolution. The results show significant temporal changes in the polarization degree (PD) across three regions of interest in the 2--8 keV energy band.
Regions a and b, located in the northern and the southwest parts of the study area, exhibit PD variations with significance levels greater than 4$\sigma$ and 3$\sigma$, respectively. Region c, located in the southwest, shows a notable decrease in PD with a significance greater than 5$\sigma$. 
However, no significant variation in the polarization angle (PA) was observed. 
Meanwhile, notable flux variations were detected, likely influenced by dynamic processes such as magnetized turbulence within the nebula. 

\end{abstract}


\section{Introduction} \label{sec:intro}

The Crab Nebula is the archetype of pulsar wind nebulae (PWNe) originated through the interaction of the ultra-relativistic wind injected by the rapidly rotating Crab pulsar (PSR B0531+21 or PSR J0534+2200), located at a distance of $\sim$2 kpc \citep{Trimble+1968AJ.....73..535T}. 
The Crab pulsar has a rotation period of 34\,ms, and the history and general properties of the system are nicely summarized in \cite{2008ARA&A..46..127H}. Optical and X-ray images of the inner nebula reveal key structures such as the inner ring, the toroidal wisps, and the jets \citep{Hester+1995,Hester+2002}. These features make the Crab Nebula a prime environment for studying the physics of compact objects, particle acceleration, and relativistic outflows \citep{Gaensler+2006ARA&A..44...17G}. 

The Crab Nebula exhibits notable flux variability across multiple wavelengths.
The 8 GHz nebula flux \citep{Aller+1985ApJ...293L..73A} showed a decrease of {0.167\% $\pm$ 0.015\% yr$^{-1}$} between 1968 and 1984, consistent with the predictions of an expanding synchrotron-emitting cloud \citep{Reynolds+1984ApJ...278..630R}. \cite{Smith+2003MNRAS.346..885S} reported a decrease in the nebula-integrated flux of {0.5\% $\pm$ 0.2\% yr$^{-1}$} at optical wavelength from 1987 to 2002.
X-ray studies have shown variations on both short and long timescales, with up to 10\% changes in flux over days to weeks \citep{Ling+2003ApJ...598..334L}, and year-scale trends observed by instruments such as Swift/BAT, INTEGRAL and BeppoSAX 
\citep{Verrecchia+2007A&A...472..705V, Wilson+2011ApJ...727L..40W}. Moreover, significant local variations in X-ray emission are also observed in the nebula \citep{Greiveldinger+1999ApJ...510..305G}. These variations are attributed to changes in shock acceleration mechanisms or magnetic field fluctuations within the nebula.

Spectral analysis revealed significant spatial variations in X-ray emission arising in different regions \citep{Mori+2004ApJ...609..186M}. From 2005 to 2008, the overall X-ray/gamma-ray emission of the Crab Nebula exhibited a decline of approximately 3.5\% per year \citep{Wilson+2011ApJ...727L..40W}. In the same period, the long-term light curve shows that the Crab goes through variations on a multi-year time scale with accompanying slope changes of a few percent \citep{Shaposhnikov+2012ApJ...757..159S}, further highlighting the nebula's dynamic nature.
In addition, multi-year-scale morphological changes in the torus and the southern jet of the nebula are also observed through the Chandra X-ray Observatory \citep{Mori+2004IAUS..218..181M}.
\cite{Pavlov+2001ApJ...554L.189P} discovered that the Vela PWN also shows temporal variability.
Similar dynamic behavior has been observed in the PWNe powered by PSR J1846-0258 \citep{Ng+2008ApJ...686..508N}, PSR B1259-63 \citep{Kargaltsev+2014ApJ...784..124K}, and PSR J0633+1746 \citep{Pavlov+2010ApJ...715...66P}.
This indicates that dynamic behavior is a common feature of PWNe.

Polarization measurements are key to understanding the magnetic field structure and the interaction of the pulsar wind with the surrounding medium. In particular, particles responsible for X-ray and $\gamma$-ray emission suffer from severe synchrotron cooling. 
The X-ray polarimeter on board OSO-8 detected significant X-ray polarization, confirming that synchrotron radiation is the dominant emission mechanism \citep{Weisskopf+1978ApJ...220L.117W}. 
Subsequent studies, including those using AstroSat CZTI \citep{Vadawale+2018NatAs...2...50V}, PoGO+ \citep{Chauvin+2018MNRAS.477L..45C}, and PolarLight \citep{fenghua+2020NatAs...4..511F}, have explored the integrated polarization properties of the nebula. More recently, the Imaging X-ray Polarimeter Explorer (IXPE) \citep{Soffitta+2021AJ....162..208S,Weisskopf+2022JATIS} provides spatially resolved polarization measurements in X-ray, allowing us to study the variations of the Crab Nebula's polarization across different regions.

Crab Nebula has been observed four times by IXPE in three years \citep[see Table \ref{tab:ixpe_observations}]{Bucciantini+2023NatAs...7..602B, Wong+2024ApJ...973..172W}, providing us with the opportunity to explore the temporal evolution of its polarization properties in greater detail. In this paper, we focus on investigating the dynamic nature of PD and PA in the Crab PWN from these four observation periods.
In Section 2, we present the IXPE observations, polarization analysis, and spectral fitting data analysis. In Section 3, we summarize the results, and in Section 4, we discuss the origin of the observed variability.

\begin{table}[ht]
\centering
\setlength{\tabcolsep}{12pt}
\caption{IXPE observations of the Crab PWN}
\label{tab:ixpe_observations}
\begin{tabular}{ccc}
\hline
\textbf{Observation ID} & \textbf{Observation Date} & \textbf{Duration (ks)} \\ \hline
01001099 & February-March, 2022 & 91 \\
02001099 & February-April, 2023 & 148 \\
02006001 & October 9-10, 2023 & 60 \\
03009601 & August 19-20, 2024 & 65 \\
\hline
\end{tabular}
\end{table}

\section{Observations and Data Reduction} \label{sec:data}

IXPE is a NASA-ASI observatory fully dedicated to measuring soft X-ray polarization \citep{Soffitta+2021AJ....162..208S, Weisskopf+2022JATIS}. The spacecraft hosts three identical grazing-incidence telescopes, providing imaging, timing, and spectral polarimetry in the 2--8 keV nominal energy band. Each telescope has a polarization-sensitive detector unit (DU), designated DU1, DU2 and DU3, equipped with a gas pixel detector \citep{Costa+2001Natur.411..662C,Baldini+2021APh,2022AJ....164..103D}. Since its launch in 2021 December, IXPE has observed several PWNe, including Crab \citep{Bucciantini+2023NatAs...7..602B, Wong+2024ApJ...973..172W}, Vela PWN \citep{Xie+2022Natur.612..658X}, MSH 15-5(2) \citep{Romani+2023ApJ...957...23R}, and PSR B0540-69 \citep{Xie+2024ApJ...962...92X}.
The field of view of IXPE is 12.$^\prime$9 $\times$ 12.$^\prime$9 square, and the angular resolution is $\sim$30$^{\prime\prime}$ \citep{Weisskopf+2022JATIS}, which is sufficient to spatially resolve the Crab PWN.

IXPE has observed Crab PWN, and the observational data are publicly archived from the High-Energy Astrophysics Science Archive Research Center (HEASARC). Data analysis is performed with IXPEOBSSIM V31.0.1 \citep{Baldini+2022SoftX}, which is developed by the IXPE Collaboration following the formalism reported in \cite{Kislat+2015APh}. This formalism is suitable for the model-independent \texttt{PCUBE} algorithm to derive polarization.

All data files were processed with the following steps before data analysis:
(1) The particle and instrumental background events were removed using the method described in \cite{DiMarco+2023AJ....165..143D}. The good time intervals filter, which excludes observational periods with a poor aspect \citep{Baldini+2022SoftX}, eliminated less than 1\% of events for each detector unit, ensuring high-quality data for further analysis. 
(2) The World Coordinate System was aligned to the pulsar position by comparing the observed and simulated Stokes I maps, with small adjustments to the right ascension and declination, significantly improving the accuracy of polarization measurements.
Additionally, the first observation included an extra pulse invariant correction. See \cite{Bucciantini+2023A&A...672A..66B,Wong+2024ApJ...973..172W} for further details of the corrections.

We identified our three elliptical regions of interest (see Figure 1) using the SAOImageDS9 software \citep{Joye+2006ASPC..351..574J}. Since the Crab PWN is extremely bright, all analyzes were performed without background subtraction \citep{DiMarco+2023AJ....165..143D}.
This analysis was performed using both the model-independent \texttt{PCUBE} algorithm in the \texttt{IXPEOBSSIM} and the \textsc{xspec} \citep{Arnaud96} unweighted spectro-polarimetric analysis \citep{2022AJ....163..170D}, with \textsc{xspec} further utilized to fit the spectrum for each observation, enabling the monitoring of flux variations over time within the regions of interest.

Spectral analysis was performed using data from the three detector units.
Considering that the selected regions are located on the outskirts of the PWN, we used the \texttt{ixpecalcarf} tool in HEAsoft version 6.34 \citep{heasoft} to generate the appropriate auxiliary response files and redistribution matrix files for each region, which were then used to calculate the flux.

\section{Results} \label{sec:Results}
\subsection{Polarization Map of the Crab}
The main objective of this analysis is to investigate whether there are significant variations in PD between different observations of the same region of the Crab Nebula. To achieve this, we compared the polarization maps of four observations and assessed the significance of the observed PD differences.

\begin{figure}
    \centering
    \includegraphics[width=\textwidth]{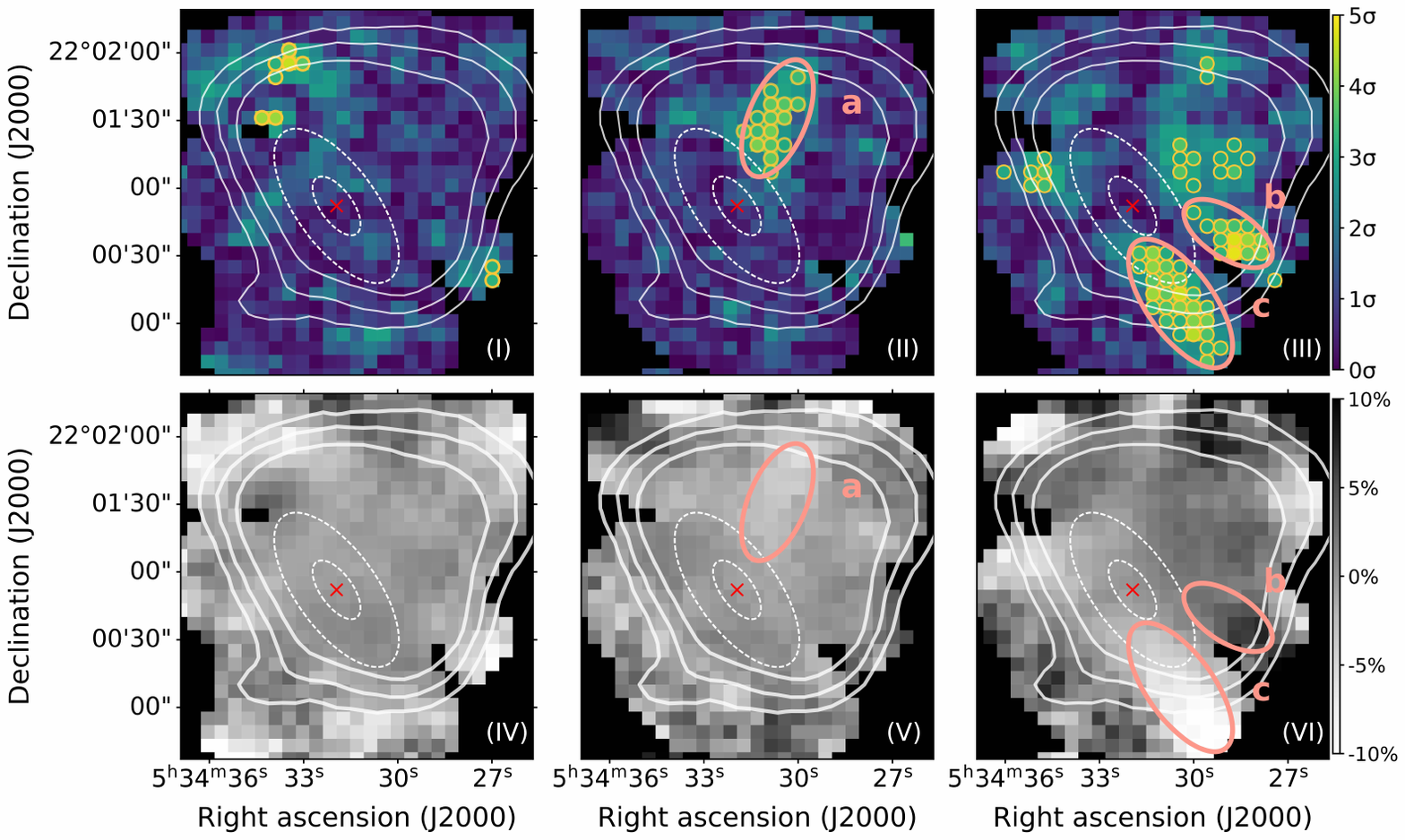}
    \caption{$\Delta$PD significance ($\tau$) maps (top) and $\Delta$PD map (bottom) of the Crab PWN based on IXPE four observations. The maps use $6^{\prime\prime}$ square pixels. 
    From left to right, the maps display deviations between the first and second observations (panel I and IV), second and third observations (panel II and V), and third and fourth observations (panel III and VI), respectively.
    The red cross is the position of the Crab pulsar.
    The solid white lines represent the Crab Nebula’s contours. The dashed lines denote the Crab’s elliptical rings, from the innermost to the outermost, corresponding to the Crab’s inner ring and torus \citep{Ng+2004ApJ...601..479N}. Yellow circles highlight regions where $\tau \geq 3$, indicating significant PD variations. The pink ellipses outline the regions of interest, named Regions a, b, and c, respectively.}  
    \label{fig:4_pd_sig_diff_map}
\end{figure}

Polarization maps were generated for each observation using the \texttt{IXPEOBSSIM} software in the 2--8 keV energy range. Event selection was carried out with the \texttt{xpselect} tool, and polarization maps were calculated using \texttt{xpbin} with the \texttt{PMAPCUBE} method \citep{Baldini+2022SoftX}. The field of view was divided into a 30×30 grid, with each grid cell being a $6^{\prime\prime}$ square bin, and a Gaussian kernel with a size of $30^{\prime\prime}$ was applied to smooth the data, guided by the mirror point spread function (PSF). We selected pixel grids with PD significance levels greater than 5 for subsequent analysis.

To quantify the significance of the variation in PD between two observations in the same region, we introduce the parameter \textit{$\tau$} = $\frac{\text{PD}_{\text{diff}}}{\varsigma}$, where \(\text{PD}_{\text{diff}}\) is the difference in PD between the two observations, and $\varsigma$= $\sqrt{\text{A}_{err}^2 + \text{B}_{err}^2}$ is the combined error, where $A_{err}$ and $B_{err}$ are the errors in PD for the two observations, respectively. A significant change in PD is defined when \textit{$\tau$} $\geq$ 3, indicating a statistically significant variation. 
The comparison of PD analysis is made on the premise that the position of the same region in the four observations is perfectly aligned. We calculated the PD variation significance (\textit{$\tau$} value) and the difference in PD ($\Delta$PD) for each $6^{\prime\prime}$ bin in two successive observations by subtracting the PD values observed earlier from those observed later in the polarization maps.

\begin{figure}
\centering
\includegraphics[width=1.0\textwidth]{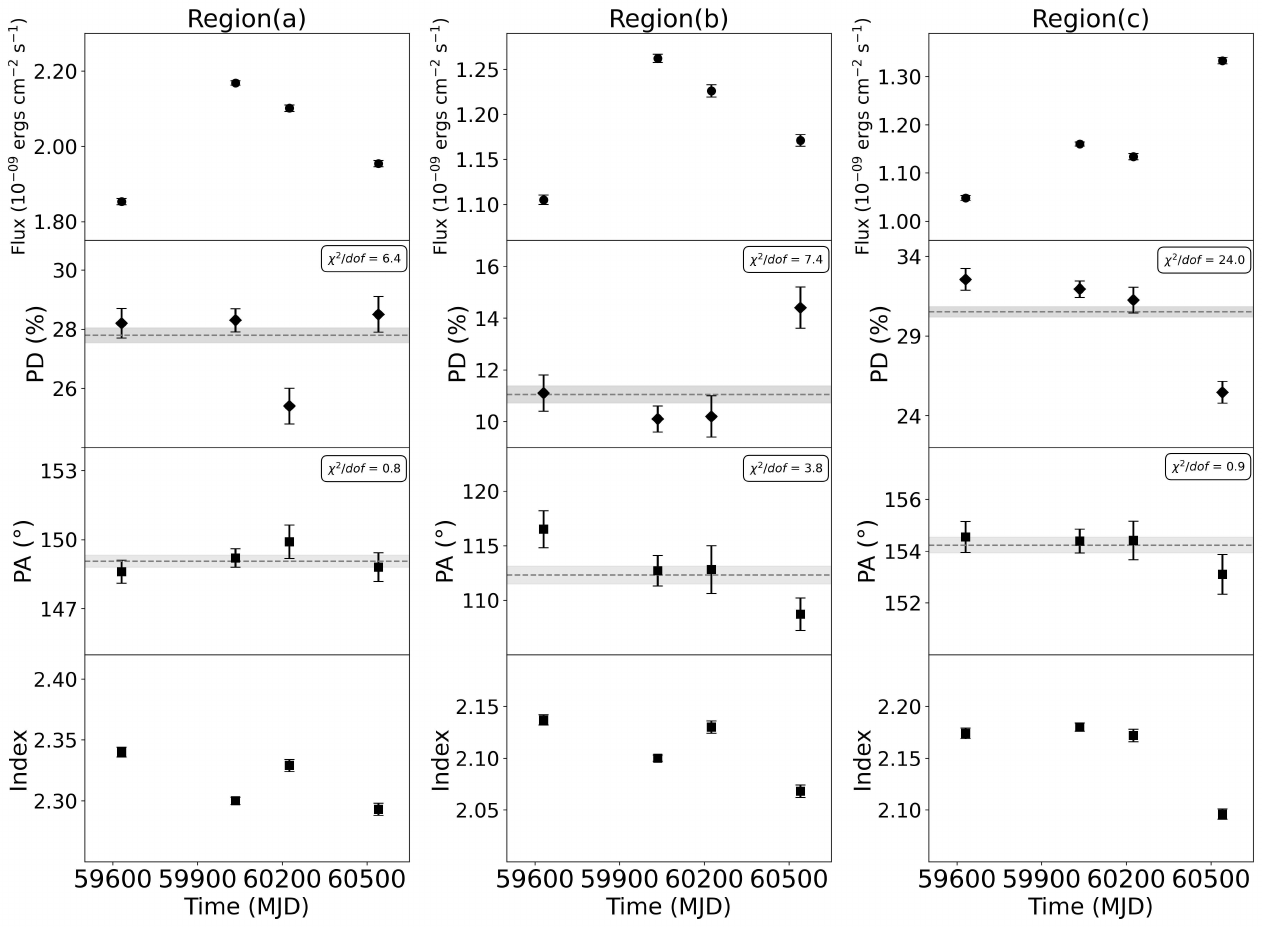}
\caption{Flux, PD, PA, and the photon index as a function of time of the Regions a, b, and c, as identified in Figure \ref{fig:4_pd_sig_diff_map}. The dashed line represents the weighted average of the PD and PA, and the gray shading indicates the 1$\sigma$ error range of the weighted average. The $\chi^2/dof$ of PD and PA relative to the weighted average value are also presented.}
\label{fig:4_flux_PD_PA}
\end{figure}

The resulting PD variation significance maps and $\Delta$PD maps, which highlight the spatial variation of PD differences, are shown in Figure~\ref{fig:4_pd_sig_diff_map}. The solid white lines trace the contours of the Crab Nebula as observed by the IXPE, while the dashed lines indicate its elliptical rings, ranging from the inner ring to the torus, with specific parameters provided by \cite{Ng+2004ApJ...601..479N}.
The top three images in Figure~\ref{fig:4_pd_sig_diff_map} show the significance maps of PD variation, and yellow circles highlight regions with significant changes ($\tau \geq 3$).
The changes between the first and second observations (panel I) are scattered at the edges of the nebula. Between the second and third observations (panel II), and between the third and fourth observations (panel III), the most significant PD changes were observed in the northern and southwestern parts of the Crab Nebula. 
Based on the concentration of yellow circles, we identified three key regions of interest, shown as pink areas on the maps denoted Region a, Region b, and Region c, for further detailed analysis.

The temporal variations in PD across these regions are evident, as shown in the $\Delta$PD maps on the bottom row of Figure \ref{fig:4_pd_sig_diff_map}, where positive values indicate an increase in PD and negative values indicate a decrease. 
The PD decreases in Region a between the second and third observations (panel V); PD increases in Region b and decreases in Region c between the third and fourth observations (panel VI). These findings highlight notable temporal variations in PD that are specific to each region.

\subsection{Polarimetric and Spectra Analysis}
We performed a more detailed analysis for the three regions of interest, with the IXPE results of flux, PD, PA, and powerlaw index shown in Figure \ref{fig:4_flux_PD_PA}, and the corresponding data provided in Table \ref{Table:Flux_PD_PA_for_Different_Regions}.
Using \texttt{xselect}, we extracted Stokes $I$, $Q$, and $U$ spectra from the IXPE Level-2 event lists. For energy binning, we used a single 2--8 keV bin. In order to apply the $\chi^2$ statistics, we require a minimum of 50 counts in each spectral channel for the I spectra. A constant 0.2 keV energy binning is then applied to the Q and U Stokes spectra. The fitting model we used is \texttt{TBabs*powerlaw}, and the spectral fitting was performed in \textsc{xspec}. The column density $N_H$ is fixed at $0.36 \times 10^{22} \text{ cm}^{-2}$ \citep{Ge+2012ApJS..199...32G}. Then we froze the parameters of the normalization model to the values previously obtained and used the \texttt{TBabs*cflux*powerlaw} model to ultimately calculate the unabsorbed flux.
We compare the flux changes between two observations to evaluate the variation over time.

\begin{table}[ht]
\centering
\setlength{\tabcolsep}{12pt}
\caption{Flux, PD, PA,  photon index, $\chi^2/dof$, and $\tau$ for the regions of interest.}
\begin{tabular}{ccccccccc}
\toprule
\textbf{Reg.} & \textbf{Obs.} & \textbf{Flux ($10^{-9}$ ergs/cm$^2$/s)} & \textbf{Index} & \textbf{PD (\%)} & \textbf{PA ($^{\circ}$)} & \textbf{$\chi^2$/dof} & \textbf{$\tau$}$^{1}$ \\ 
\midrule
\multirow{4}{*}{\textbf{a}} 
& 1 & $1.854 \pm 0.008$ & $2.339 \pm 0.004$ & $28.2 \pm 0.5$ & $148.6 \pm 0.5$ & 652/409 & 0.7 \\ 
& 2 & $2.168 \pm 0.006$ & $2.301 \pm 0.003$ & $28.3 \pm 0.4$ & $149.2 \pm 0.4$ & 463/428 & 1.1 \\ 
& 3 & $2.101 \pm 0.009$ & $2.329 \pm 0.008$ & $25.4 \pm 0.6$ & $149.8 \pm 0.7$ & 440/372 & 3.7 \\ 
& 4 & $1.954 \pm 0.009$ & $2.293 \pm 0.007$ & $28.5 \pm 0.6$ & $148.8 \pm 0.6$ & 423/395 & 0.9 \\ 
\midrule
\multirow{4}{*}{\textbf{b}} 
& 1 & $1.105 \pm 0.005$ & $2.137 \pm 0.005$ & $11.1 \pm 0.7$ & $116.5 \pm 1.7$ & 451/391 & 0.1 \\ 
& 2 & $1.262 \pm 0.005$ & $2.100 \pm 0.004$ & $10.1 \pm 0.5$ & $112.7 \pm 1.4$ & 403/416 & 1.6 \\ 
& 3 & $1.226 \pm 0.007$ & $2.130 \pm 0.006$ & $10.3 \pm 0.8$ & $112.8 \pm 2.2$ & 418/371 & 1.0 \\ 
& 4 & $1.171 \pm 0.006$ & $2.068 \pm 0.006$ & $14.4 \pm 0.8$ & $108.7 \pm 1.5$ & 429/378 & 3.9 \\ 
\midrule
\multirow{4}{*}{\textbf{c}} 
& 1 & $1.048 \pm 0.005$ & $2.174 \pm 0.005$ & $32.5 \pm 0.7$ & $154.5 \pm 0.5$ & 386/391 & 2.7 \\ 
& 2 & $1.161 \pm 0.004$ & $2.179 \pm 0.004$ & $31.8 \pm 0.5$ & $154.4 \pm 0.5$ & 464/397 & 2.3 \\ 
& 3 & $1.135 \pm 0.007$ & $2.172 \pm 0.007$ & $31.2 \pm 0.8$ & $154.4 \pm 0.7$ & 396/340 & 0.9 \\ 
& 4 & $1.333 \pm 0.007$ & $2.100 \pm 0.005$ & $25.5 \pm 0.7$ & $153.0 \pm 0.8$ & 412/392 & 6.6 \\  
\bottomrule
\end{tabular}
\vspace{0.3cm}
\raggedright \textsuperscript{1} Calculated  from the difference between the PD of each region and the average PD. 
\label{Table:Flux_PD_PA_for_Different_Regions}
\end{table}

Flux variations across the three regions of interest revealed distinct trends during the four observation periods. Between the first and second observations, the flux increased by more than 10\% in all three regions. In the period from the second to the third observations, the flux consistently declined by approximately 3\% across all the three regions. From the third to the fourth observations, Regions a and b exhibited further declines, with flux decreasing by around 7\% and 4\%, respectively. In contrast, Region c demonstrated a significant increase in flux, increasing by 17.6\% $\pm$ 0.7\%. 
During the fourth observation period, the unique upward trend in flux for Region c demonstrated significant differences in the level of dynamism across different areas within the PWN.
In addition, Chandra images have shown filamentary structures and small-scale morphology changes over time; these dynamic processes may also contribute to the observed evolution of flux over time.

We performed spectropolarimetric fitting, using both \texttt{IXPEOBSSIM} and \textsc{xspec} for cross-validation, to ensure the consistency of the polarization results across the three regions. 
Based on \texttt{constant*TBabs*polconst*powerlaw} model, the polarization model \texttt{polconst} in \textsc{xspec} was used to fit the Stokes $I$, $Q$, and $U$ spectra within the 2--8 keV band, extracting polarization information.
The model-independent analysis was performed using \texttt{IXPEOBSSIM}, while the fitting model was applied in \textsc{xspec}. The resulting PA and PD obtained from \texttt{IXPEOBSSIM} and \textsc{xspec} were found to be largely consistent, as shown in Table \ref{tab:ixpeobssim_xspec}.
Significant variations in PD are visible in Figure \ref{fig:4_flux_PD_PA} row two. 
We calculated the significance of PD variations (\textit{$\tau$} values) in three regions based on the PD from the adjacent previous and subsequent observations. In Region a, we observed two substantial changes in PD, both exceeding $4\sigma$. The PD decreased during the third observation and returned to its original level by the fourth observation. 
In Region b, between the third and fourth observations, PD increased with significance above $3\sigma$. However, in Region C, the PD significantly decreased by more than $5\sigma$.

To verify the variation of PD and PA, we used the weighted average values as a constant model and calculated the $\chi^2/dof$ for each region. In Figure \ref{fig:4_flux_PD_PA}, this is shown with a dashed line, and the gray shaded area represents the 1$\sigma$ error range. We estimated the deviations of PD and PA from the weighted averages, and the $\chi^2/dof$ for each region are marked in the corresponded panel.
The PD values in all three regions exhibit relatively high $\chi^2/dof$, suggesting that a constant cannot adequately fit the data.
Then, the \textit{$\tau$} on PD for each observation was computed based on these average values, as shown in Table \ref{Table:Flux_PD_PA_for_Different_Regions}.
In Region c, the significance of the fourth observation of PD reaches $6.6\sigma$.
The particularly high significance of PD variations observed in Region c indicates the dynamic evolution of the degree of local magnetic field ordering in the Crab PWN over time.

\begin{table}[h!]
\centering
\setlength{\tabcolsep}{18pt}

    \caption{Comparison of polarization parameters across four observations in regions of interest using IXPEOBSSIM and \textsc{xspec}}
    \renewcommand{\arraystretch}{1.2} 
    \begin{tabular}{lccc}
        \toprule
        PD & Region a & Region b & Region c \\
        \midrule
        PD1\textsuperscript{I} & 28.2$\pm$0.5 & 11.1$\pm$0.7 & 32.5$\pm$0.7 \\
        PD1\textsuperscript{X} & 29.6$\pm$0.5 & 11.8$\pm$0.6 & 31.5$\pm$0.7 \\
        PA1\textsuperscript{I} & -31.4$\pm$0.5 & -63.5$\pm$1.7 & -25.5$\pm$0.6 \\
        PA1\textsuperscript{X} & -32.4$\pm$0.5 & -64.5$\pm$1.6 & -25.7$\pm$0.6 \\
        \midrule
        PD2\textsuperscript{I} & 28.3$\pm$0.4 & 10.1$\pm$0.5 & 31.8$\pm$0.5 \\
        PD2\textsuperscript{X} & 29.9$\pm$0.4 & 11.3$\pm$0.5 & 31.8$\pm$0.5 \\
        PA2\textsuperscript{I} & -30.8$\pm$0.4 & -67.3$\pm$1.4 & -25.7$\pm$0.5 \\
        PA2\textsuperscript{X} & -32.1$\pm$0.5 & -66.1$\pm$1.2 & -26.7$\pm$0.4 \\
        \midrule
        PD3\textsuperscript{I} & 25.4$\pm$0.6 & 10.3$\pm$0.8 & 31.2$\pm$0.8 \\
        PD3\textsuperscript{X} & 27.4$\pm$0.6 & 10.9$\pm$0.8 & 31.2$\pm$0.8 \\
        PA3\textsuperscript{I} & -30.2$\pm$0.7 & -67.2$\pm$2.2 & -25.6$\pm$0.7 \\
        PA3\textsuperscript{X} & -31.6$\pm$0.6 & -66.3$\pm$2.0 & -26.3$\pm$0.7 \\
        \midrule
        PD4\textsuperscript{I} & 28.5$\pm$0.6 & 14.4$\pm$0.8 & 25.5$\pm$0.7 \\
        PD4\textsuperscript{X} & 29.3$\pm$0.6 & 15.3$\pm$0.7 & 25.9$\pm$0.6 \\
        PA4\textsuperscript{I} & -31.2$\pm$0.6 & -71.3$\pm$1.5 & -27.0$\pm$0.8 \\
        PA4\textsuperscript{X} & -31.6$\pm$0.5 & -70.8$\pm$1.4 & -27.7$\pm$0.7 \\
        \bottomrule
    \end{tabular}
    \label{tab:ixpeobssim_xspec}
    \vspace{0.5em}
    \begin{minipage}{\textwidth}
        \raggedright \textsuperscript{I} Values are obtained with IXPEOBSSIM. \\
        \raggedright \textsuperscript{X} Values are obtained with XSPEC.
    \end{minipage}
\end{table}

\begin{figure}
\centering
\includegraphics[width=0.6\textwidth]{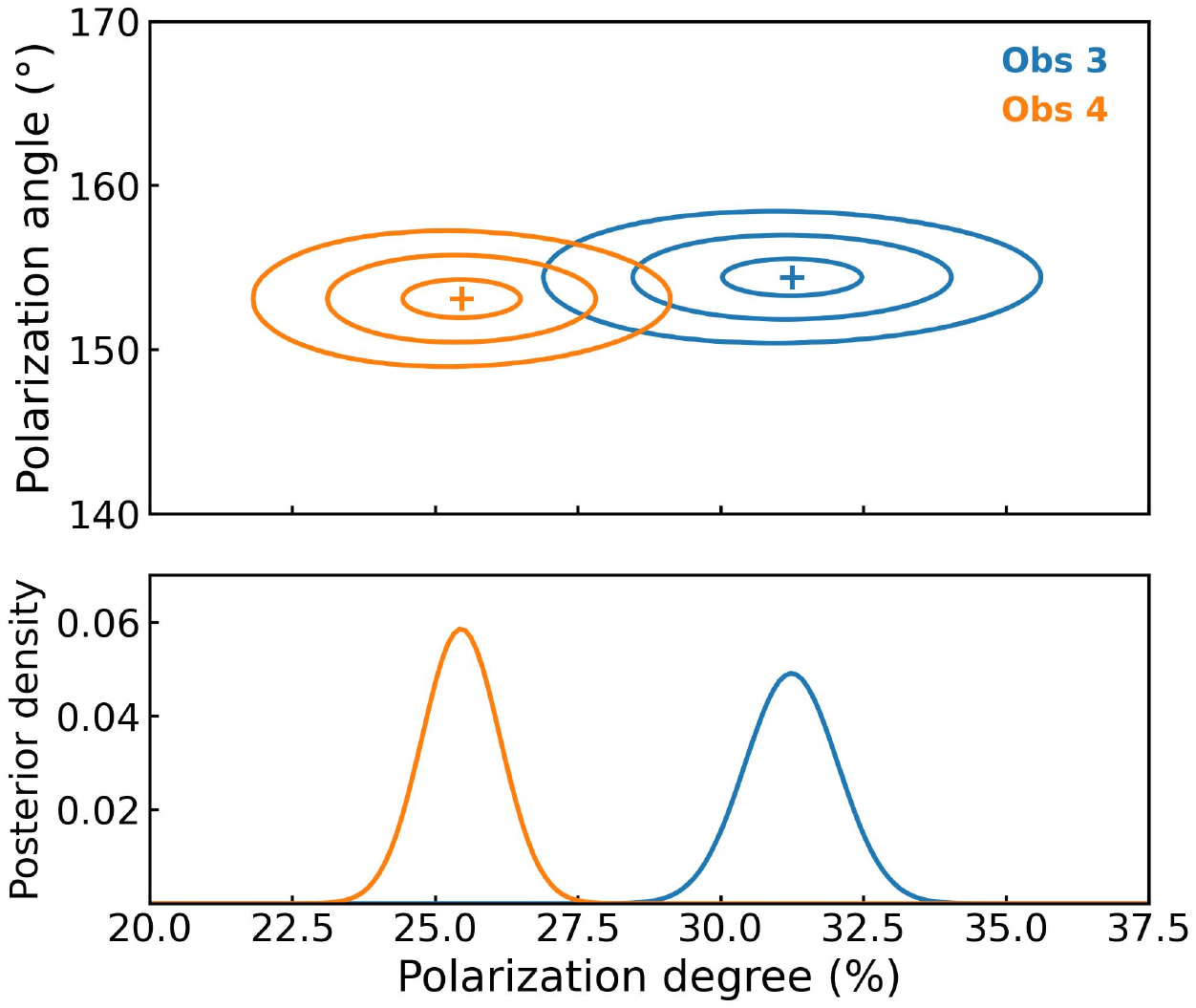}
\caption{Top: The posterior distributions of PD and PA for Observation 3 (orange) and Observation 4 (blue) in Region c. The contours enclose the 1$\sigma$, 3$\sigma$, and 5$\sigma$ levels. Bottom: The posterior distribution of PD. Each measurement is inconsistent at a level 3.4$\sigma$.}
\label{fig:Posterior distributions}
\end{figure}

We also employed a Bayesian approach to assess changes in polarization parameters, following the methodology outlined in \cite{Maier+2014PASP..126..459M}. Here, we compare two samples from Region c, using polarization data from the third and fourth observations. 
The posterior distributions of polarization for the two samples are plotted in Figure \ref{fig:Posterior distributions}. Each measurement is inconsistent with the other at a level of 3.4$\sigma$.

Furthermore, PA variations across the three regions were found to be essentially constant over the four observations. The $\chi^2/dof$ values for Region a and Region c are approximately 1, indicating that the constant fits well to the PA values. We also notice that the $\chi^2/dof$ value in Region b is slightly higher than 1, while
the $\tau$ values for PA in both the first and fourth observations relative to the weighted average are both less than 3. Therefore, we conclude that PA remains unchanged, suggesting that the large-scale magnetic field structure is stable and does not evolve over time, while magnetic turbulence might cause the variation of PD over time.

\section{Discussion and Conclusions} \label{sec:Discussion}

\begin{figure}
\centering
\includegraphics[width=0.6\textwidth]{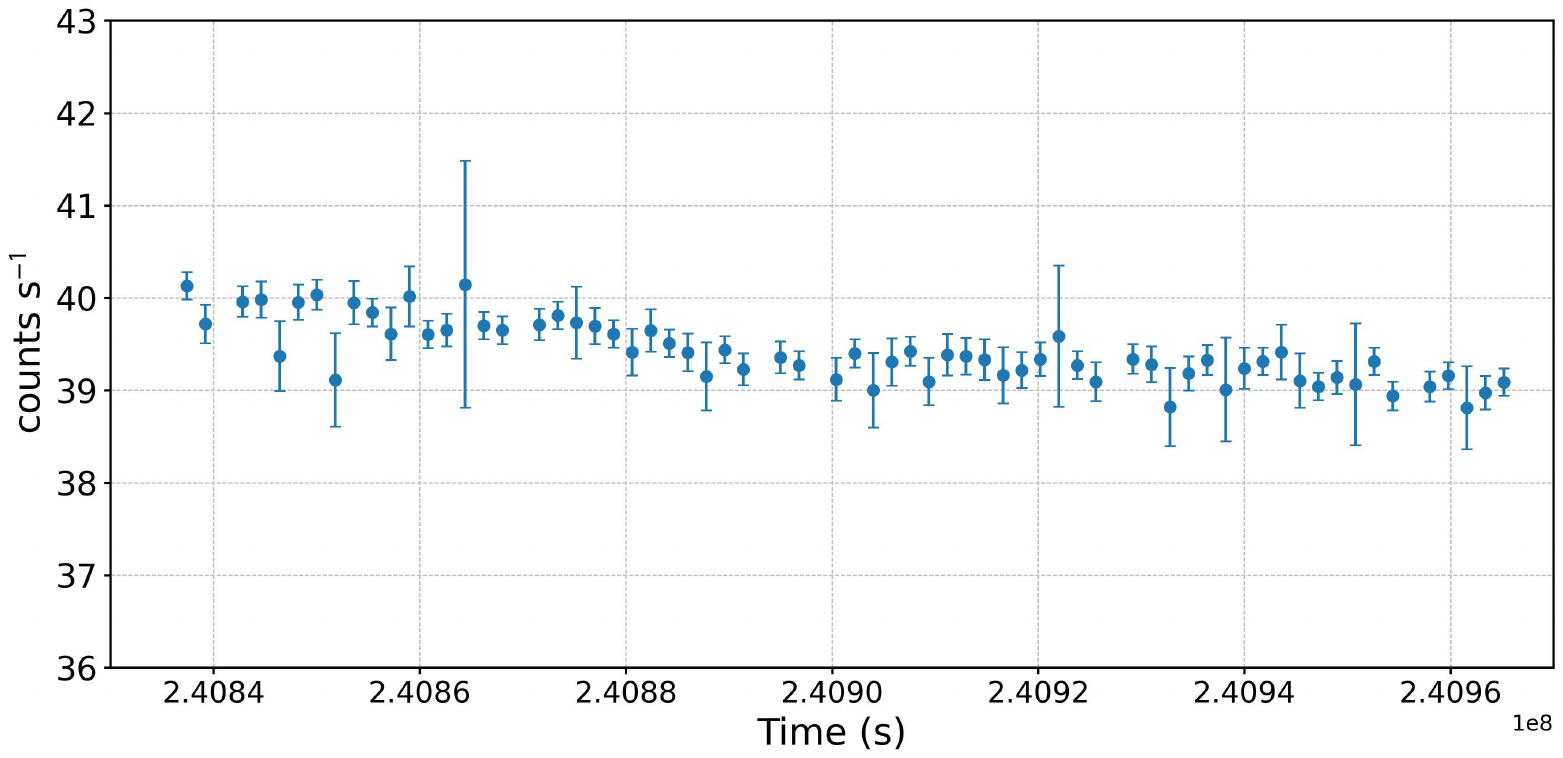}
\caption{The light curve of the Crab Nebula from the fourth observational data set, covering the entire IXPE field of view. The analysis was performed in the 2--8 keV energy band with a 0.5 hour time bin using the three DUs.}
\label{fig4:light curve}
\end{figure}

This study used IXPE data to investigate the relationship between flux, PD, and PA in the Crab Nebula. 
In all three regions of interest (defined in Figure \ref{fig:4_pd_sig_diff_map}), flux initially increased and then decreased, with an anomalous flux increase observed in Region c during the fourth observation, see Figure \ref{fig:4_flux_PD_PA}.
We checked the light curve of the Crab Nebula from the fourth observation, which was generated without applying any region selection, encompassing the entire IXPE field of view, as shown in Figure \ref{fig4:light curve}.
This light curve used data from three DUs in the 2--8 keV energy band, and the time bin is 0.5 hour.
The overall light curve revealed no significant fluctuations.
Therefore, we ruled out the influence of solar activity on the flux increase, suggesting that the flux variation in this region is likely due to intrinsic changes within the nebula.

During the first observation, the Crab pulsar was about $2.8^{\prime}$ off-axis relative to the mirror system \citep{Bucciantini+2023NatAs...7..602B}. This misalignment may import uncertainty in the Image Response Functions, which are essential for precise spectral analysis. Consequently, the flux measurements from this observation are likely unreliable and should be interpreted with caution.

One possible reason for the variation in X-ray PD could be magnetized turbulence. Magnetized turbulence produces reconnecting current sheets of various sizes throughout the bulk of the nebula. Particles are then accelerated by magnetic reconnection in the current layers and by scattering off turbulent fluctuations \citep{Luo+2020ApJ...896..147L}. This is a highly dynamic process where turbulence enhances the randomness of the magnetic field, and these turbulent interactions contribute to the temporal changes in PD observed. However, a predominant field direction can still be identified, and therefore the PA in each region remains constant.
Magnetized turbulence also leads to spatial inhomogeneity in the magnetic field, affecting synchrotron emission characteristics and potentially driving the flux variations.
However, \cite{Wilson+2011ApJ...727L..40W} attributed flux changes to shock acceleration or variations in the nebular magnetic field. On the basis of the current data, we analyze the relationship between photon index, PD, and synchrotron emission efficiency, and no clear correlations were observed in the localized regions of the Crab Nebula.

We estimate the upper limit of the energy ratio between the random and ordered magnetic fields using the method from \cite{Bandiera+2016MNRAS.459..178B} and \cite{Bucciantini+2017MNRAS.470.4066B} in Region a. This region was specifically selected because the magnetic field curvature in this region is very small, remaining nearly parallel. In contrast, Regions b and c exhibit significantly higher curvature, which can lead to severe depolarization effects, causing turbulence evaluations in these areas to be inaccurate.
The observed PD for a local region can be defined as:

\[
\Pi = \frac{\alpha + 1}{\alpha + 5/3}
\times \frac{3 + \alpha}{4}
\times \frac{\sin^2 \theta_B}{2 \sigma^2}
\times \frac{_1F_1\left( \left(1-\alpha\right)/2, 3; -\sin^2 \theta_B / 2 \sigma^2 \right)}{_1F_1\left( -\left(1+\alpha\right)/2, 1; -\sin^2 \theta_B / 2 \sigma^2 \right)}
\]
Here, $\alpha$ is the spectral index, $\theta_{B}$ is the angle between the magnetic field direction and the line of sight in the comoving coordinate.
$\mathit{{}_1F_1}(a, b; x)$ is the Kummer confluent hypergeometric function, with a and b representing the upper and lower parameters, and $\mathit{x}$ being the argument of the function. $\sigma = \sqrt{E_r / 3E_o} = B_r / \sqrt{3}B_o$ is related to the energy ratio between the random and ordered magnetic fields, assuming a three-dimensional isotropic Gaussian random field with a variance of \((B\sigma)^2\) in each direction.
Here, $B_r$, $B_o$, and $B$ refer to the strengths of the random, ordered, and total magnetic fields, respectively. The photon index ($\Gamma$) for Region a is measured at 2.328, determined using the \texttt{TBabs*powerlaw} model fit, so the spectral index $\alpha$ is 1.328 ($\Gamma$ = $\alpha$ + 1). With a PD of 25.4\% and under extreme conditions $\theta_{B} \sim 90^\circ$, the maximum of $B_r/B_o$ is 1.8, corresponding to a magnetic energy ratio $E_r/E_o$ of 3.3 in Region a. 

During the IXPE observations of the Crab, no simultaneous Chandra ACIS observations were conducted. However, significant spatial and temporal variations have been reported in the soft X-ray band \citep{Greiveldinger+1999ApJ...510..305G,Hester+2002ApJ...577L..49H,Mori+2004ApJ...609..186M}, implying multicomponent variable spectra. Filamentary structures seen in Chandra images \citep{Hester+2002ApJ...577L..49H,Mori+2004IAUS..218..181M} suggest that the magnetic fields are different according to locations in the Crab Nebula. Furthermore, annual-scale morphological changes in both the torus and the southern jet have been observed \citep{Mori+2004IAUS..218..181M}. The PWN exhibits smaller-scale structures that are highly dynamic, varying over short timescales and different energy emissions. These IXPE observations emphasize the Crab Nebula's intrinsic spatial and temporal variability, with changes in flux, PD, and structural transformations over time, while the stable PA indicates that the large-scale magnetic field structure remains constant.

In addition to the three regions studied, we found other relatively smaller areas of the Crab Nebula also showed significant changes in both flux and PD, while the PA variation remained constant across the four observations. However, the sizes of them are much smaller than the PSF of IXPE, thus they are not disclosed in detail here. This indicates that the observed phenomena are more widespread throughout the nebula. The consistent changes across multiple regions suggest that the nebula's dynamic nature drives these variations, supporting the hypothesis that such changes are consistent with the presence of dynamic processes, like turbulence, which could contribute to the observed variations in flux, PD, and PA.

Future research should focus on extended temporal monitoring, including observations made every six months, to further explore the dynamic interactions between flux, PD, and PA. 
Continuous observations with advanced polarization instruments will provide deeper insights into the complex magnetic field structures and their evolution in the Crab Nebula and similar astrophysical objects.

\section*{Acknowledgments}
This work is supported by National Key R\&D Program of China (grant No. 2023YFE0117200), and National Natural Science Foundation of China (grant No. 12373041 and 12422306), and Bagui Scholars Program (XF). This work is also supported by the Guangxi Talent Program (“Highland of Innovation Talents”). 
This research used data products provided by the IXPE Team (MSFC, SSDC, INAF, and INFN) and distributed with additional software tools by the High-Energy Astrophysics Science Archive Research Center (HEASARC), at NASA Goddard Space Flight Center (GSFC). The Imaging X-ray Polarimetry Explorer (IXPE) is a joint US and Italian mission.  
ADM and FLM contribution is supported by the Italian Space Agency (Agenzia Spaziale Italiana, ASI) through contract ASI-INAF-2022-19-HH.0, and by the Istituto Nazionale di Astrofisica (INAF) in Italy. 
ADM and FLM are partially supported by MAECI with grant CN24GR08 “GRBAXP: Guangxi-Rome Bilateral Agreement for X-ray Polarimetry in Astrophysics”.


\end{document}